\documentclass[12pt,titlepage]{article}
\usepackage[english]{babel}
\usepackage[dvips]{graphicx}

\def\BR{\rm BR}

\begin{document}
\begin{titlepage}
%23 April
\vspace{.5cm}

\begin{center}
{\Large \bf Charm Radiative Decays with Neutral Mesons
 $D^0 \to \bar K^0 \pi^0 \gamma,\,D^0 \to \bar K^0 \eta(\eta^\prime) \gamma$\\}

\vspace{.5cm}

{\large \bf S. Fajfer$^{a,b}$, A. Prapotnik$^{b}$ and
P. Singer$^{c}$\\}

\vspace{0.5cm}
{\it a) Department of Physics, University of Ljubljana,
Jadranska 19, 1000 Ljubljana, Slovenia}\vspace{.5cm} 

{\it b) J. Stefan Institute, Jamova 39, P. O. Box 300, 1001 Ljubljana,
Slovenia}\vspace{.5cm}

{\it c) Department of Physics, Technion - Israel Institute  of Technology,
Haifa 32000, Israel\\}

\end{center}

\vspace{1.5cm}

\centerline{\large \bf ABSTRACT}

\vspace{0.5cm}

 The radiative decays $D^0 \to \bar K^0 P^0 \gamma$ with nonresonant
$\bar K^0 P^0$ ($P^0= \pi^0,\eta,\eta^\prime$) are considered within the 
framework which combines heavy quark effective theory and the chiral 
Lagrangian. Due to neutral mesons the amplitudes do not have bremsstrahlung 
contributions. We assume factorization for the weak matrix elements. Light 
(virtual) vector mesons are found to give the main contribution to the decay 
amplitudes. The decay $D^0 \to \bar K^0 \pi^0 \gamma$ is predicted to have a 
branching ratio of $3\times 10^{-4}$, with comparable contributions from 
parity-conserving and parity-violating parts of the amplitude. 
The decays with $\eta (\eta^\prime)$ in the
final state are expected with branching ratios of $1.1\times 10^{-5}$ and
$0.4\times10^{-7}$ respectively and are mainly parity conserving.

\end{titlepage}

\setlength {\baselineskip}{0.55truecm}
\setcounter{footnote}{1}    % start footnotes at dagger instead
			     % of '*' (article style only)

\setcounter{footnote}{0}

Radiative weak decays of mesons have proved to be a fertile ground for the
investigation of models of the strong interactions involved, and of the
nonleptonic weak Hamiltonian. This is true for long-distance dominated 
K decays \cite{1} as well as for B radiative decays, driven by short-distance
dynamics [2-4]. One is lead to expect that likewise, radiative decays of D
mesons [5-10] will bring similar useful insights, as well as providing
possible checks [11-15] for physics beyond the standard model.

 In a recent paper \cite{16}, we have initiated the study of
radiative decays of type $D \to K \pi \gamma$, with non resonant $K \pi$. 
In \cite{16} we studied the Cabibbo allowed decays which have a bremsstrahlung
component, $D^+ \to \bar  K^0 \pi^+ \gamma$ and $D^0 \to  K^- \pi^+ \gamma$.
The theoretical framework used is that of the effective Lagrangian which
contains both heavy flavor symmetry and $SU(3)_L \times SU(3)_R$ chiral
symmetry \cite{17} and the calculation was carried out with the factorization 
approximation for the weak matrix elements. The merging of this framework
with the known QED requirements for the bremsstrahlung radiation
necessitated the development of a specific technique for the treatment 
\cite{16} of these decays.

  In the present article we undertake the study of decays with neutral mesons
 $D^0 \to \bar K^0 \pi^0(\eta,\eta^\prime)\gamma$, which are likewise Cabibbo 
allowed but do not involve a bremsstrahlung component. Due to this feature, 
these purely direct radiative decays may provide a cleaner testing ground for 
the study of the dynamics involved and for exploring the suitability of the
theoretical framework employed in their calculation. In \cite{16} it was found
that the branching ratio for the parity-conserving part of the decay
$D^0 \to K^- \pi^+ \gamma$, which part is of purely "direct" nature, can be 
as high as $10^{-4}$, while the total branching ratio for photon energies 
above 50~MeV is close to $10^{-3}$. Such relatively high rates, especially 
for the direct transition, strongly motivates the desirability of studying the
decays with neutral mesons which we consider here.

 In decays of this type, one does not expect any significant short-distance 
contribution [5-10]. Since the decays $D^0 \to \bar K^0 P^0 \gamma$
involve transitions between a rather heavy meson and pseudoscalars,
we adopt for our calculation the heavy quark chiral Lagrangian ($HQ\chi L$)
containing heavy flavor and $SU(3)_L \times SU(3)_R$ symmetries as the 
appropriate theoretical framework. However, since virtual vector mesons may 
play an important role in these decays in the ($P \gamma$) channels, we need 
to complement the Lagrangian \cite{17} with the light vector mesons. For this 
purpose we use the generalization of $HQ\chi L$ by Casalbuoni et al. \cite{18}
to include the vector mesons in the Lagrangian, in which the original symmetry
is now broken spontaneously to diagonal $SU(3)_V$ \cite{19}. This framework is 
described in detail in Refs. \cite{8,18}, henceforth we recapitulate only the
main features. The light meson part of the Lagrangian is written as
\begin{eqnarray}
\label{deflight}
{\cal L}_{light} & = & -{f^2 \over 2}
\{tr({\cal A}_\mu {\cal A}^\mu) +
a\, tr[({\cal V}_\mu - {\hat \rho}_\mu)^2]\}\nonumber\\
& + & {1 \over 2  g_v^2} tr[F_{\mu \nu}({\hat \rho})
F^{\mu \nu}({\hat \rho})]\;.
\end{eqnarray}
with $a=2$ for exact vector meson dominance \cite{19}. The vector and axial-vector
currents are given by
\begin{eqnarray}
\label{VA}
{\cal V}_{\mu} =  \frac{1}{2} (u^{\dag} D_{\mu} u + u D_{\mu}u^{\dag}) \qquad
{\cal A}_{\mu}  =  \frac{1}{2} (u^{\dag}D_{\mu} u - u D_{\mu}u^{\dag})\;,
\end{eqnarray}
where $u=\exp  \left( \frac{i \Pi}{f} \right)$, $\Pi$ being the matrix of 
the pseudoscalar fields and $f=f_\pi=132$~MeV. $D_\mu$ is the covariant 
derivative. Moreover, ${\hat \rho}_\mu=i {g_v \over \sqrt{2}} \rho_\mu$,
with $\rho_\mu$ being the matrix of the vector fields:  
\begin{eqnarray}
\label{defrho}
\rho_\mu = \pmatrix{
{\rho^0_\mu + \omega_\mu \over \sqrt{2}} & \rho^+_\mu & K^{*+}_\mu \cr
\rho^-_\mu & {-\rho^0_\mu + \omega_\mu \over \sqrt{2}} & K^{*0}_\mu \cr
K^{*-}_\mu & {\bar K^{*0}}_\mu & \Phi_\mu \cr}.
\end{eqnarray}
For the vector coupling $g_v(m_V^2)$ we use experimentally determined values 
from leptonic decays \cite{20} of vector mesons, which accounts for the 
symmetry breaking effects.

In the heavy sector, to order $O(p)$ one has the Lagrangian
\begin{eqnarray}
\label{HQ}
{\cal L}_{str} &=& Tr(H_aiv \cdot D_{ab} \bar H_b)+
igTr(\bar H_a H_b \gamma_\mu A^\mu_{ba}\gamma_5),  
\end{eqnarray}
where $D_{ab}^\mu H_b = \partial^\mu H_a - H_b V_{ba}^\mu$, while the trace 
$Tr$ runs over Dirac indexes. The flavor indexes are denoted as a and b.
Both the heavy  
pseudoscalar and the heavy vector meson are incorporated in a 4x4 matrix
\begin{eqnarray}
\label{defHM}
 H_a=\frac{1}{2}(1+ \!\!\not{\!} v)(P_{\mu a}\gamma^\mu-P_{5a}\gamma_5)
\end{eqnarray}
The strong coupling constant has been determined \cite{21} recently from
$D^* \to D \pi$ decay to be $g \approx 0.59$ and $v_\mu$ is the four-velocity 
of D meson.

The electromagnetic interaction is introduced by gauging the Lagrangians
(\ref{deflight}) and (\ref{HQ}) with the U(1) photon field, thus amending 
appropriately the covariant derivatives. However, the gauging procedure alone 
does not generate the $D^* \to D \gamma$ transition, thus requiring the 
introduction of additional terms ${\cal L}_c$. There are two such terms
in the framework used, giving the direct photon-heavy quark interaction with 
strength $\lambda^{\prime}$ (being of the order $1/m_Q$) and a light vector 
meson-heavy meson interaction with strength $\lambda$ (being of the order 
$1/\lambda_{\chi}$ where $\lambda_{\chi}$ is the chiral perturbation theory
scale). The second term provides photon emission via vector meson dominance 
(VMD). 
Thus,
\begin{eqnarray}
\label{defoddheavy}
{\cal L}_{c} & = & - {\lambda}^{\prime} Tr [H_{a}\sigma_{\mu \nu}
F^{\mu \nu} (B) {\bar H_{a}}] + i {\lambda} Tr [H_{a}\sigma_{\mu \nu}
F^{\mu \nu} (\hat \rho)_{ab} {\bar H_{b}}].  
\end{eqnarray}
An additional contributing term to the radiative decays via VMD is the
Wess-Zumino-Witten anomalous interaction for the light sector, given 
by \cite{22}:
\begin{eqnarray}
\label{defloddlight}
{\cal L}^{(1)}_{odd} & = & -4 \frac{C_{VV\Pi}}{f} \epsilon
^{\mu \nu \alpha \beta}Tr (\partial_{\mu}
{\rho}_{\nu} \partial_{\alpha}{\rho}_{\beta} \Pi).
\end{eqnarray}
In the calculation, instead of using the SU(3) symmetric coupling, we
shall rely on the experimentally measured effective couplings of the 
$V \to P \gamma$ transitions \cite{23}. For $\lambda$, $\lambda^\prime$ we take 
$\lambda=-0.49$~$GeV^{-1}$, $\lambda^\prime=-0.102$~$GeV^{-1}$, as determined 
from $D^{*+,0}$ electromagnetic and strong decays \cite{16}, the signs chosen as to 
conform  with the quark model.
The effective weak $\Delta c=1$ nonleptonic Lagrangian of relevance to
the decays investigated here is
\begin{eqnarray}
\label{deflfermisl}
{\cal L}_{NL}^{eff}(\Delta c=\Delta s=1)
& = & -{G_F \over \sqrt{2}} V_{ud}V_{cs}^*
[ a_1 O_1 +  a_2 O_2 ],
\end{eqnarray}
where $O_1=(\bar s c)^\mu_{V-A}(\bar u d)_{\mu,V-A}$, 
$O_2=(\bar u c)^\mu_{V-A}(\bar s d)_{\mu,V-A}$ and $V_{ij}$ 
are the CKM matrix elements. 
The effective Wilson coefficients are taken as $a_1=1.26$, $a_2=-0.55$ 
\cite{24}. 
In our calculation we rely on factorization, which implies 
$<P^0 \bar K^0|O_1|D^0>=0$. The heavy-light weak current is bosonized as 
\cite{17,18}
\begin{eqnarray}
\label{jqbig}
{J_Q}_{a}^{\mu} = &\frac{1}{2}& i \alpha Tr [\gamma^{\mu}
(1 - \gamma_{5})H_{b}u_{ba}^{\dag}]\nonumber\\
&+& \alpha_{1}  Tr [\gamma_{5} H_{b} ({\hat \rho}^{\mu}
- {\cal V}^{\mu})_{bc} u_{ca}^{\dag}]\nonumber\\
&+&\alpha_{2} Tr[\gamma^{\mu}\gamma_{5} H_{b} v_{\alpha}
({\hat \rho}^{\alpha}-{\cal V}^{\alpha})_{bc}u_{ca}^{\dag}]+...\;,
\end{eqnarray}
The constant $\alpha$ is then related to the decay constant $f_D$ by
$\alpha=f_D \sqrt{m_D}$, while $\alpha_1$, $\alpha_2$ are determined from the
values of the form factors appearing in $D \to V l \nu$ decays, as explained 
in \cite{16}. Their numerical values are then 
$|\alpha_1|=0.156$~${\rm GeV}^{1/2}$,
$|\alpha_2|=0.052$~${\rm GeV}^{1/2}$.

The general Lorentz decomposition of the $D^0 \to \bar K^0 P^0 \gamma$
amplitude is given by
\begin{eqnarray}
\label{matrix}
{\cal M}=-\frac{G_F}{\sqrt{2}} V_{du} V^*_{cs} \left( F_1 \left( 
(q \cdot \varepsilon)(p \cdot k)-(p \cdot \varepsilon) (q \cdot k)\right)
+F_2 \epsilon^{\mu \alpha \beta \gamma} \varepsilon_\mu v_\alpha k_\beta 
q_\gamma\right)\;,
\end{eqnarray}
where $F_1$, $F_2$ are the electric and magnetic transitions, which are
respectively parity-violating and parity-conserving. The four-momenta  
of $P^0$, $K$ and the photon are $q$, $p$, $k$  respectively and
$\varepsilon_\mu$ is the polarization vector of the photon.

\begin{figure}[thb]
\includegraphics[width=15cm]{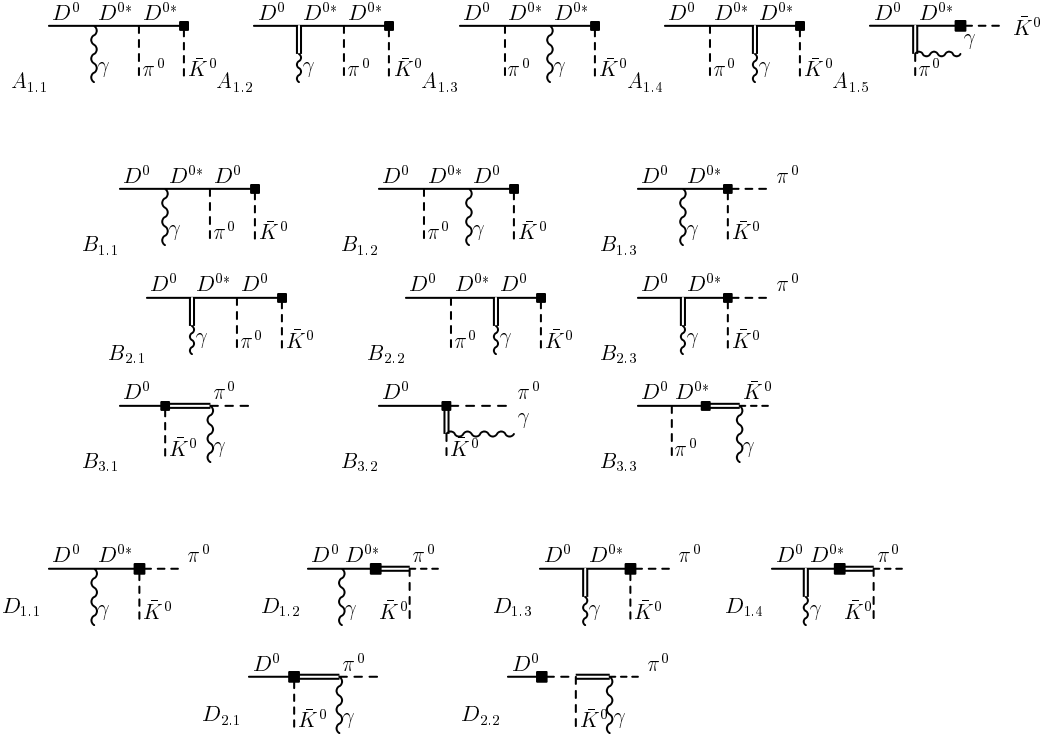}
\caption{Feynman diagrams}
\end{figure}

In Fig.1 we exhibit the Feynman diagrams contributing to the $D^0 \to
\bar K^0 \pi^0 \gamma$ decay. The diagrams denoted $A_i$ (parity-violating)
derive from the $<\pi^0|J|D^0><\bar K^0|J|0>$ term; the same term
gives the parity-conserving terms $B_i$, while the parity-conserving terms
$D_i$ come from $<0|J|D^0><\pi^0 \bar K^0|J|0>$. The amplitudes $A_i$, $B_i$, 
$D_i$ are related to the $F_1$, $F_2$ form factors by
\begin{equation}
\label{F1}
F_1(D^0 \to \bar K^0 P^0 \gamma) = A_1\;,
\end{equation}
\begin{equation}
\label{F2}
F_2(D^0 \to \bar K^0 P^0 \gamma) = \sum_{i=1}^3 B_i+ \sum_{i=1}^2 D_i\;. 
\end{equation}
The sums for each row of the amplitudes in Fig.1 are gauge invariant. The
explicit expressions of $A_i$, $B_i$, $D_i$ are listed below, for the
$D^0 \to \bar K^0 \pi^0 \gamma$ decay.

$$A_1=i\frac{\sqrt{2}}{2}eg\frac{f_{D} f_K}{f_\pi}
\frac{v \cdot k}{v \cdot k+v \cdot q+\Delta}
\left(\frac{1}{v \cdot k+\Delta}-\frac{1}{v \cdot q+\Delta}\right)
\left(2\lambda^{'}-\frac{\sqrt{2}}{2}\lambda g_v\left(\frac{q_\omega}
{3m_\omega^2}+\frac{q_\rho}{m_\rho^2}\right)\right)$$
\begin{equation}
\label{A}
-ief_{D} f_K g_v\lambda M \frac{(v \cdot k)(p \cdot k)^2}
{(v \cdot (q+k)+\Delta)}
\left(\frac{g_{\rho \pi \gamma}}{(q+k)^2-m_{\rho}^2+i\Gamma_{\rho}m_{\rho}}+
\frac{g_{\omega \pi \gamma}}{(q+k)^2-m_{\omega}^2+i\Gamma_{\omega}m_{\omega}}
\right)\;,
\end{equation}

\vspace{0.5cm}

$$B_{1}=-\sqrt{2}eM\frac{f_{D} f_K}{f_\pi}\lambda^{,}\left(\frac{1}
{v \cdot k+\Delta}+g\frac{v \cdot p}{(v \cdot q+v \cdot k)(v \cdot k+\Delta)}+
g\frac{v \cdot p}{(v \cdot q+v \cdot k)(v \cdot q+\Delta)}\right)\;,$$
$$B_{2}=\frac{1}{2}Me\frac{f_D f_K}{f_\pi} \lambda g_v \left(\frac{q_\omega}
{3m_\omega^2}+\frac{q_\rho}{m_\rho^2}\right) \left(\frac{1}{v \cdot k+\Delta}+
g\frac{v \cdot p}{(v \cdot q+v \cdot k)(v \cdot k+\Delta)}\right.$$
\begin{equation}
\label{B}
\left. +g\frac{v \cdot p}{(v \cdot q+v \cdot k)(v \cdot q+\Delta)}\right)\;,
\end{equation}
$$B_{3}=\sqrt{M}eg_v f_K (\alpha_1 M-\alpha_2 v \cdot p)
\left(\frac{g_{\rho \pi \gamma}}{(q+k)^2-m_{\rho}^2+i\Gamma_{\rho}m_{\rho}}+
\frac{g_{\omega \pi \gamma}}{(q+k)^2-m_{\omega}^2+i\Gamma_{\omega}
m_{\omega}}\right)$$
$$-\frac{\sqrt{2}}{2}eg_{\bar K^{0*} \bar K^0 \gamma}g_{K^*} 
\frac{f_D}{f_\pi}M\frac{1+g\frac{M-v \cdot q}{v \cdot q+\Delta}}
{(k+p)^2-m_{K^*}^2+i\Gamma_{K^*}m_{K^*}}\;,$$

\vspace{0.5cm}

$$D_{1}=\frac{\sqrt{2}}{2}Me f_D \frac{1}{v \cdot k+\Delta}
\left(1+ \frac{m_{K^*}^2}{(p+q)^2-m^2_{K^*}+im_{K^*}\Gamma_{K^*}}\right)
\left(2\lambda^{'}-\frac{\sqrt{2}}{2}\lambda g_v\left(\frac{q_\omega}
{3m_\omega^2}+\frac{q_\rho}{m_\rho^2}\right) \right)\;,$$
\begin{equation}
\label{D}
D_{2}=M\frac{\sqrt{2}}{2}e\frac{f_D}{f_\pi}\left(\frac{g_{\rho}
g_{\rho \pi \gamma}}{(q+k)^2-m^2_{\rho}+im_{\rho}\Gamma_{\rho}}
-\frac{g_{\omega}g_{\omega \pi \gamma}}{(q+k)^2-m^2_{\omega}+
im_{\omega}\Gamma_{\omega}}\right)
\end{equation}
$$-\frac{\sqrt{2}}{4}ef_D f_K \frac{M^3}{(M^2-m^2_{\bar K^0})}
\left(\frac{m^2_{\rho}}{g_{\rho}}\frac{g_{\rho \pi \gamma}}
{(q+k)^2-m^2_{\rho}+im_{\rho}\Gamma_{\rho}}-
\frac{m^2_\omega}{g_\omega}\frac{g_{\omega \pi\gamma}}
{(q+k)^2-m^2_{\omega}+im_{\omega}\Gamma_\omega}\right)\;.$$
\noindent

  The treatment of the decays to $\eta$, $\eta^\prime$ requires the inclusion 
of $\eta-\eta^\prime$ mixing. This is usually performed \cite{25} by the 
mixing of singlet and octet states via a unitary matrix characterized by an 
angle $\theta$. The analysis of various decays involving $\eta$, $\eta^\prime$
and of meson masses has led to a mixing angle in the range ($-9^\circ$) to 
($-23^\circ$). However, a treatment \cite{26} based on chiral perturbation 
theory and a detailed phenomenological analysis indicates that this simple 
scheme is inadequate and must be extended \cite{26} to include two mixing 
angles $\theta_0$ and $\theta_8$, in addition to the two decay constants.
Here we follow this approach with 
$|\eta>=\cos \theta_8|\eta_8>-\sin \theta_0|\eta_0>$ and  
$|\eta^\prime>=\sin \theta_8|\eta_8>+\cos \theta_0|\eta_0>$.
For the $D^0 \to \bar K^0 \eta \gamma$ and $D^0 \to \bar K^0 \eta^{\prime} 
\gamma$ amplitudes the same form as Eqs. (\ref{A}-\ref{D}) holds, except 
for replacement of constants.
In order to obtain these amplitudes one has to replace in above Eqs.
$f_\pi$ by $f_\eta/(K^\eta_d\sqrt{2})$ or 
$f_{\eta^\prime}/(K^{\eta^\prime}_d\sqrt{2})$,
while in amplitude $D_1$ one replaces
$\left(1+ \frac{m_{K^*}^2}{(p+q)^2-m^2_{K^*}+im_{K^*}\Gamma_{K^*}}\right)$ by
$\sqrt{2}\left(K^{\eta}_d+(K^{\eta}_d-K^{\eta}_s)
\frac{m_{K^*}^2}{(p+q)^2-m^2_{K^*}}\right)$.
Moreover, one replaces $g_{\rho\pi\gamma}$, $g_{\omega\pi\gamma}$ by
$g_{\rho\eta(\eta')\gamma}$ and $g_{\omega\eta(\eta')\gamma}$.
The factors $K^\eta_d$, $K^\eta_s$, $K^{\eta^\prime}_d$, 
$K^{\eta^\prime}_s$ are octet-singlet mixing factors, given by
\begin{equation}
K^\eta_d=\frac{\cos\theta_8}{\sqrt{6}}-\frac{\sin\theta_0}{\sqrt{3}}\;,
\qquad
K^{\eta'}_d=\frac{\sin\theta_8}{\sqrt{6}}+\frac{\cos\theta_0}{\sqrt{3}}\;,
\end{equation}
\begin{equation}
K^\eta_s=-\frac{2\cos\theta_8}{\sqrt{6}}-\frac{\sin\theta_0}{\sqrt{3}}\;,
\qquad
K^{\eta'}_s=-\frac{2\sin\theta_8}{\sqrt{6}}+\frac{\cos\theta_0}{\sqrt{3}}\;.
\end{equation}

 A recent analysis by Feldmann, Kroll and Stech 
\cite{28} has given further clarification of the theoretical basis of this 
scheme. Following their procedure \cite{27,28}, we use 
$\theta_8= -20.2^\circ$, $\theta_0=-9.2^\circ$, $f_\eta=f_\pi$ 
and $f_\eta'=1.13f_\pi$. We have used these values in our calculation, as well
as checking the sensitivity of the results to the scheme and to the values
of above constants within an acceptable range.

 Using now Eqs. (\ref{A}-\ref{D}) we calculate the differential decay
distributions and total decay rates for $D^0 \to \bar K^0 \pi^0 \gamma$, and
with appropriate replacements as indicated above, for $D^0 \to K^0 
\eta(\eta^\prime)\gamma$. Since $\rho$ is a wide resonance we use a momentum 
dependent width
\begin{equation}
\Gamma_\rho(q^2)=\Gamma_\rho(m^2_\rho)
\left(\frac{q^2-4m^2_\pi}{m^2_\rho-4m^2_\pi} \right)^{3/2}
\frac{m_\rho}{\sqrt{q^2}}\Theta(4m^2_\pi)\;.
\end{equation} 
%20
 We are interested in decays to nonresonant $K\pi$ final states and
we also wish to delete from the final state resonant ($K\gamma$) and
($P\gamma$) configurations. Accordingly, we subtract the calculated rate 
given by the diagrams that include above configurations from the total rate. 
However, our results include the contributions of the remaining virtual vector
mesons and their interference with the resonant ones. Our resulting 
prediction for the major radiative decay is
\begin{equation}
\label{21}
\BR(D^0 \to \bar K^0 \pi^0 \gamma)_{NR}=3.0 \times 10^{-4}\;.       
\end{equation}
%20 
 The parity-conserving and the parity-violating parts of the
amplitude contribute approximately 2/3 and 1/3 to the decay rate
respectively. Now, if we do not subtract the direct vector mesons 
contribution, one gets a branching ratio of $3.0 \times 10^{-3}$ for 
this channel. Obviously, the two cases have  rather different Dalitz 
plots, the resonances being easily identified in it as they dominate the 
decay rate in the latter case. On the other hand, if we exclude vector mesons
from our Lagrangian, the total decay rate is reduced to $2.6 \times 10^{-8}$.
We have also calculated the size of the rate coming from the direct 
resonant process 
$D^0 \to \bar K^{*0} \gamma$ using our formalism and we find it to be more 
than two orders of magnitude smaller than (\ref{21}), in general agreement
with previous estimates \cite{5,8,9}.

  For the decays to $\eta$, $\eta^\prime$, again after deleting resonant two
body channels from the final state, we get:
\begin{equation}
 \BR(D^0 \to \bar K^0 \eta \gamma)_{NR} = 1.1 \times 10^{-5}\;,    
\end{equation}
\begin{equation}
 \BR(D^0 \to \bar K^0 \eta^\prime \gamma)_{NR} = 4.3 \times 10^{-8}\;.
\end{equation}
%21, 22

 Allowing also $\eta\gamma$, $\eta^\prime\gamma$ in the final
state, coming from the intermediate $D^0 \to \rho (\omega) \gamma$ 
decay, these figures are 
raised to $3.9\times 10^{-5}$ and $1.4\times 10^{-7}$
respectively. Excluding completely the contribution of vector mesons, these
branching ratios are lowered to $3.7\times 10^{-8}$ and $1.3\times 10^{-9}$. 
In the decays to $\eta$, $\eta^\prime$, the dominant contribution to the decay 
rate is due to the parity conserving part of the amplitude.

  The major role played by off-the-mass-shell vector mesons in these decays is
evident from the above procedure. In particular, the $\omega$-meson exchange 
is the dominant contribution to the pionic decay (Eq.(20)), while for the decay
$D^0 \to \bar K^0 \eta \gamma$, both $\omega$ and $K^*$ exchange are giving 
major contributions. For the decay to $\eta^\prime$, the main contribution is 
due to $K^*$. The contribution of the $\rho$ meson is generally much smaller,
in all channels.

 We have also found that the $D^0 \to K \eta \gamma$ rate is smaller than 
$D^0 \to K \pi^0 \gamma$ rate partly due to phase space, but mostly due 
to the different values of the coupling constants involved. The smallness of 
the decay with $\eta^\prime$ in the final state is mainly due to phase space 
inhibition.

In Fig.2 we display the Dalitz plots for the three decay channels, and the
photon spectra of these decays. The Dalitz plot coordinates are 
$m_{12}=\sqrt{(P-k)^2}$ on x axis and  $m_{23}=\sqrt{(P-p)^2}$ on y axis
and the spectrum is expressed in terms of $(m_{12})^2=(M - k)^2$. In the
main channel $D^0 \to \bar K^0 \pi^0 \gamma$ the peak of the parity-conserving
contribution is at $E_\gamma=700$~MeV and of the parity-violating contribution
at $E_\gamma=500$~MeV. The total spectrum is predicted to peak around 
$E_\gamma=650$~MeV. For the $\eta$ decay, the spectrum peaks at 
$E_\gamma=300$~MeV.

  Before concluding we wish to make some observations concerning the
appropriateness of the theoretical framework we used. By using the
Lagrangians (\ref{HQ}), (\ref{defoddheavy})-(\ref{deflfermisl})
 as well as the factorization approximation
we can calculate also various decays of type $D \to VP$. In particular, 
we have calculated the rates for four channels of this type, which relate 
to the radiative decays we considered. Of these, $\omega$ channel is the
most important one for the pionic decay while for the $\eta$ channel, both the
$\omega$ and $K^*$ channels are giving the main contribution.
We present now the results we
obtained, giving in parentheses the observed branching ratios:
$\BR(D^0 \to \omega \bar K^0)=5.6\%(2.1\%)$,
$\BR(D^0 \to \rho   \bar K^0)=7\%(1.2\%)$,
$\BR(D^0 \to \pi    K^*)     =11\%(3.2\%)$,
$\BR(D^0 \to \eta   K^*)     =3.2\%(1.9\%)$.
The rates we obtain are within factor two-four  (the relevant ones are on the
lower side of this) of the experimental ones, which leads us to estimate that 
this is essentially the accuracy of our calculation, including also the 
uncertainty due to subtraction of the resonant channels from the total rate.
 We have checked the sensitivity of our results to the values of the
mixing angles and the values of $f_\eta$, $f_\eta'$. Within reasonable 
values for these constants, the changes in rates are negligible.

We conclude by pointing out that the branching ratio of 
$D^0\to \bar K^0 \pi^0 \gamma$ is quite large, due to the contribution of
light virtual vector mesons. That makes it rather appealing for the 
experimental study since in this decay mode there is no need for special 
treatment of the bremsstrahlung component. 

  With the abundance of D's produced at B factories, at Tevatron and 
at the forthcoming charm factories, we look forward to experimental
results for these decays, especially for the $\pi$ and $\eta$ modes,
which are predicted to have sizable rates of $3\times10^{-4}$ and 
$1.1\times 10^{-5}$ respectively.

\vspace{0.3cm}
{\bf Acknowledgments}

The research of S.F. and A.P. was supported in part by the Ministry of 
education, Science and Sport of the Republic of Slovenia. The research of P.S.
was supported in part by Fund of Promotion of Research at the Technion.

\begin{figure}
\begin{center}
\includegraphics[width=6cm]{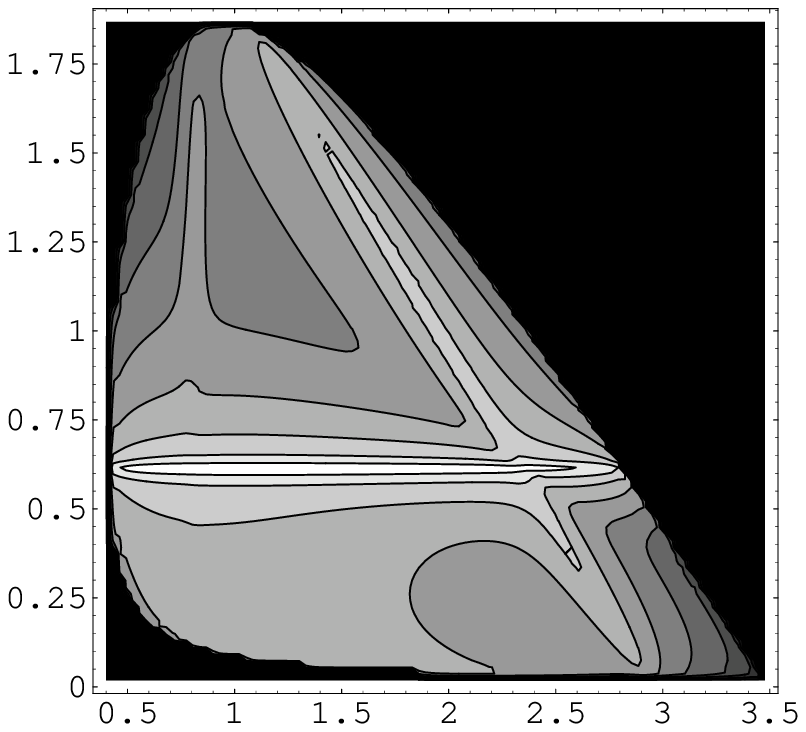}
\includegraphics[width=9cm]{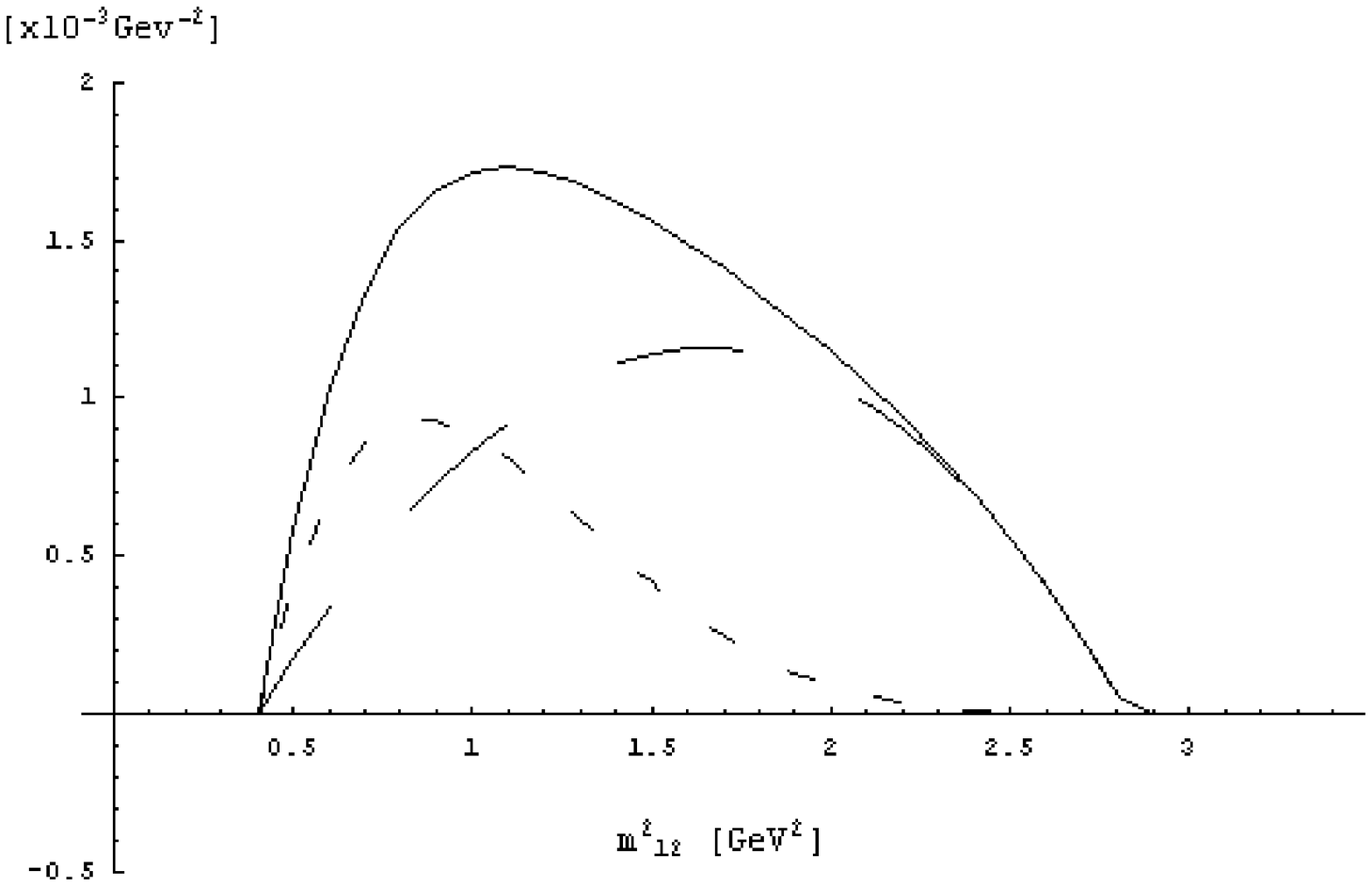}
\includegraphics[width=6cm]{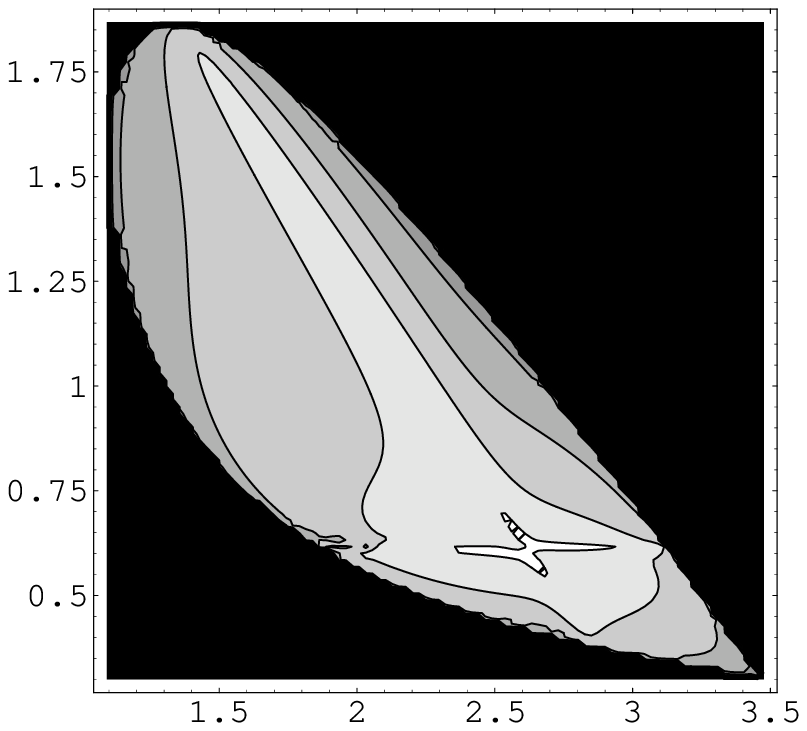}
\includegraphics[width=9cm]{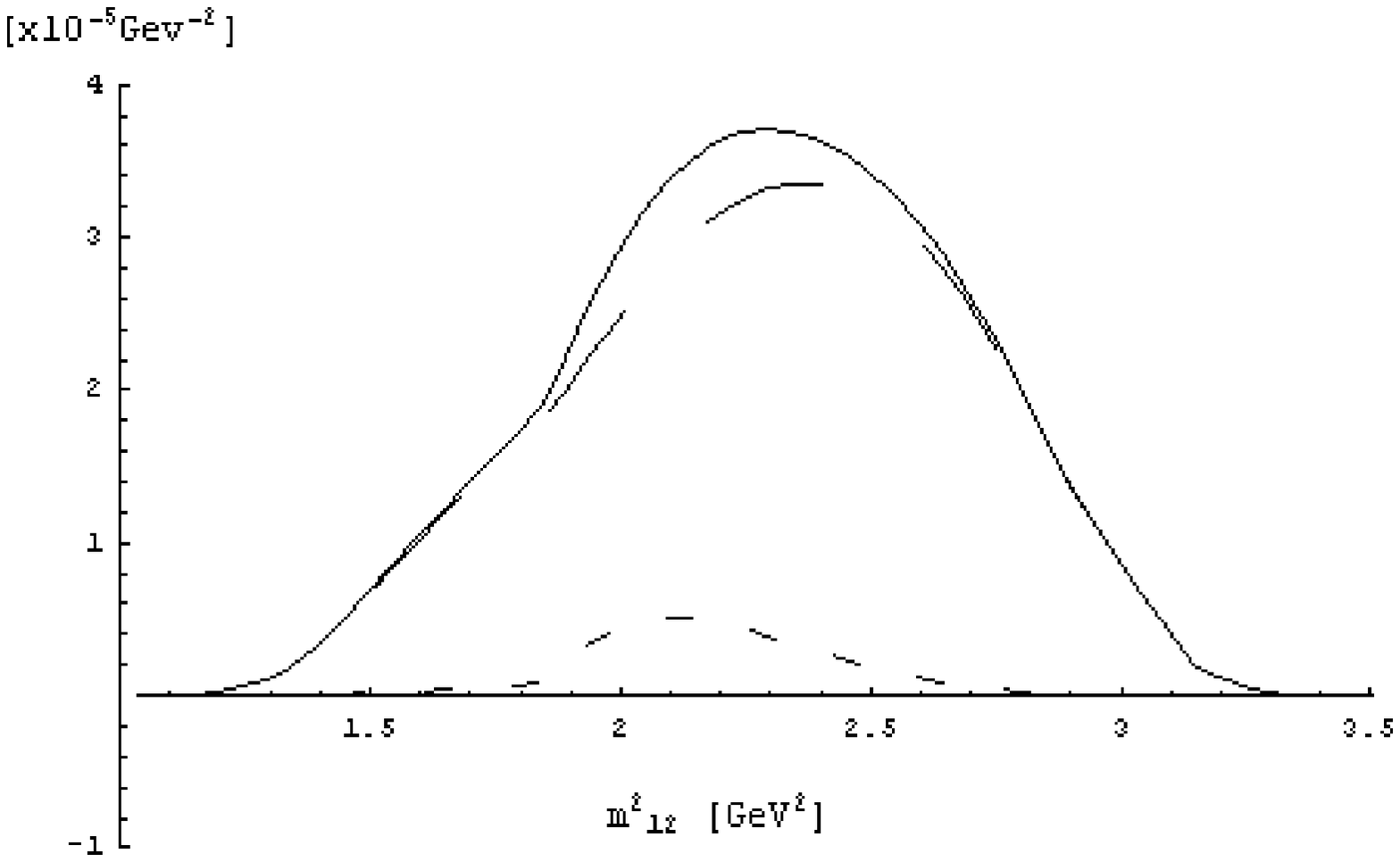}
\includegraphics[width=6cm]{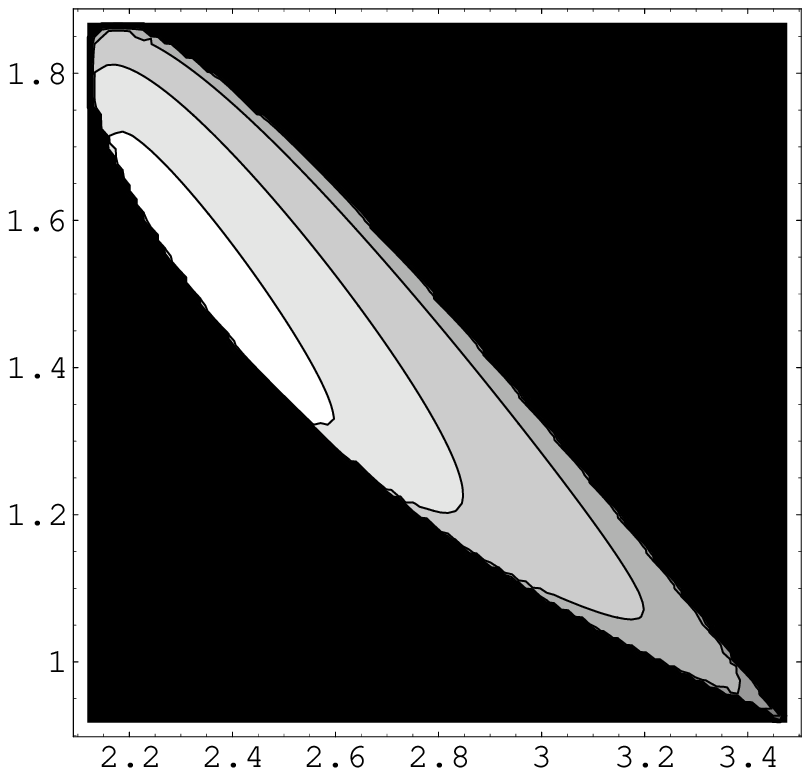}
\includegraphics[width=9cm]{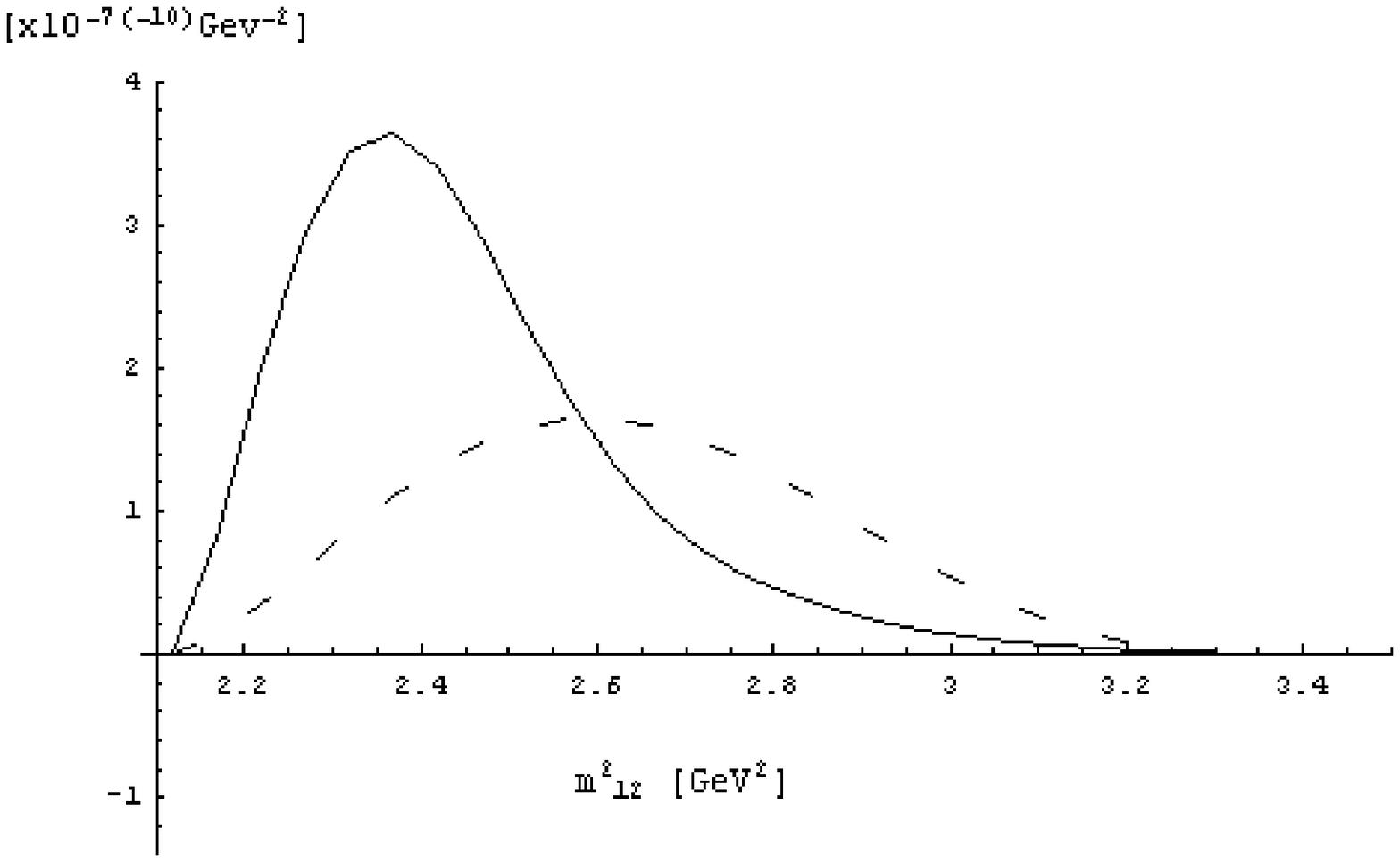}
\end{center}
\caption{Dalitz plots (1. column) and photon spectra (2. column) for 
the $D^0 \to \bar K^0 \pi^0 \gamma$ decay
(1. row), $D^0 \to \bar K^0 \eta \gamma$ decay (2. row) and 
$D^0 \to \bar K^0 \eta' \gamma$ decay (3. row). Short-dashed line: parity violating part. 
Long-dashed line: parity conserving part. Full-line: Total contribution. On the bottom right
figure $10^{-7}$ label is for the full line and $10^{-10}$ for the dashed line.}
\end{figure}

\end{document}